# Chunked-and-Averaged Estimators for Vector Parameters

Hien D. Nguyen and Geoffrey J. McLachlan[1]

**Abstract**

A divide-and-conquer method for parameter estimation is the chunked-and-averaged (CA) estimator. CA estimators have been studied for univariate parameters under independent and identically distributed (IID) sampling. We study the CA estimators of vector parameters and under non-IID sampling.

*Keywords:* asymptotic statistics; Big Data; divide-and-conquer; distributed computing; parameter estimation.

## 1. Introduction

The computation of Big Data statistics has become a common problem in the modern world. Such computations have been made difficult due to the numerous idiosyncrasies and pathologies of Big Data (cf. Fan et al., 2014). Two of the most pervasive of the pathologies of Big Data are the distribution of the data and the infeasibility of computing statistics on the entirety of the data, simultaneously.

A resolution to these pathologies have been the adoption of divide-and-conquer methodologies, which have been the traditionally dependable approach in computer science for such problems (cf. Dasgupta et al., 2008, Ch. 2). The popularity of the divide-and-conquer approach to computation is clear when considering the widespread use of **MapReduce** packages such as **Hadoop**; see for example Srinivasa and Muppalla (2015). Successful implementations of divide-and-conquer strategy in the **R** programming environment (R Core Team, 2016) are the **RHIPE** and **partools** packages; see Guha et al. (2012) and Matloff (2016c), respectively. Although divide-and-conquer provides a feasible framework for the computation of Big Data statistics, it does not in itself provide theoretical guarantees of the computed statistics.

Let $\{\boldsymbol{x}_i\}_{i=1}^n$ be the realization of the random sample $\{\boldsymbol{X}_i\}_{i=1}^n$ of size $n \in \mathbb{N}$, where $\boldsymbol{X}_i \in \mathbb{R}^d$ for $d \in \mathbb{N}$ and $i \in [n]$ ($[n] = \{1, 2, ..., n\}$). Suppose that there are two whole numbers $c_n$ and $r_n$ such that we can rewrite the realization

---

[1]HDN is at the Department of Mathematics and Statistics, La Trobe University, Melbourne. GJM is at the School of Mathematics and Physics, University of Queensland, St. Lucia. (*Corresponding author email: h.nguyen5@latrobe.edu.au)

as

$$\begin{array}{cccc} \boldsymbol{x}_{11}, & \boldsymbol{x}_{12}, & \ldots, & \boldsymbol{x}_{1c_n}, \\ \boldsymbol{x}_{21}, & \boldsymbol{x}_{22}, & \ldots, & \boldsymbol{x}_{2c_n}, \\ \ldots, & \ldots, & \ldots, & \ldots, \\ \boldsymbol{x}_{r_n1}, & \boldsymbol{x}_{r_n2}, & \ldots, & \boldsymbol{x}_{r_nc_n}, \end{array} \quad (1)$$

where $\boldsymbol{x}_{ij} = \boldsymbol{x}_{(i-1)c_n+j}$. Note that $n = r_n c_n$. We call each column of (1) a chunk, and thus the data is partitioned into $c_n$ chunks.

Suppose that the random sample arises from some unknown population $F$ and let $\boldsymbol{\theta}(F) \in \mathbb{R}^p$ be some interesting vector parameter. Let $\hat{\boldsymbol{\theta}}_{jn}(\boldsymbol{X}_{1j}, ..., \boldsymbol{X}_{r_nj})$ be an estimator of $\boldsymbol{\theta}(F)$ computed on chunk $j \in [c_n]$. We shall write the $j$th chunk estimator and the parameter as $\hat{\boldsymbol{\theta}}_{jn}$ and $\boldsymbol{\theta}$, for brevity. The chunked-and-averaged (CA) estimator is then defined as $\bar{\boldsymbol{\theta}}_n = c_n^{-1} \sum_{j=1}^{c_n} \hat{\boldsymbol{\theta}}_{jn}$.

The CA estimator provides a means of combining estimates of $\boldsymbol{\theta}$ on data distributed over $c_n$ locations, or statistics that are infeasible to compute on the entire data of size $n$. Furthermore, even when there is no need for distribution or partitioning, the CA estimator with $c_n$ chunks can offer a speedup when compared to the single-chunk estimator of $\boldsymbol{\theta}$. That is, for fixed $n$, if $\bar{\boldsymbol{\theta}}_n$ can be computed in time $O(n^a)$ for some $a \geq 1$ when $c_n = 1$, then $\bar{\boldsymbol{\theta}}_n$ can be computed in time $O(n^a/c_n^a)$ when $c_n > 1$, in parallel. Further, if only one processing unit is used, then we still obtain a speedup since the effective computation time will be $c_n$ times a process with time $O(n^a/c_n^a)$, or $O(n^a/c_n^{a-1})$ (cf. Matloff, 2016c). Here we use Landau's big-$O$ notation; see Dasgupta et al. (2008, Sec. 0.3).

In Matloff (2016c), it was shown that in the $d = 1$ case with fixed $c_n = c$ and IID (independent and identically distributed) sampling, if the chunked estimators are all asymptotically normal as $r_n \to \infty$, then the CA estimator is asymptotically normal with the same limiting mean and variance. The $d > 1$ case is suggested in Matloff (2016a) and Matloff (2016b), but was not proved. Similarly, for $d = 1$ and IID sampling, Li et al. (2013) showed that when the mean and variance of the chunked estimators exists, the asymptotic distribution of the CA estimator is normal in both the cases of $r_n = r$ fixed with $c_n \to \infty$, and when both $r_n \to \infty$ and $c_n \to \infty$.

In this article, we extend the results of Matloff (2016c) and Li et al. (2013) to the vector-parameter case of $d > 1$. Further, we provide results whereupon statistical guarantees can be given under non-IID random sampling. Both Matloff (2016c) and Li et al. (2013) note the usefulness of extending the results to non-IID data.

Examples of theoretical analyses for CA estimators in the literature include those in Guha et al. (2012) and Fan et al. (2007) who consider the specific case of a CA estimator in linear regression. Here, CA estimators are used to generate single-coefficient CIs (confidence intervals) and conduct inference on regression coefficients, one at a time. Our results can be used to extend upon the work of Fan et al. (2007) in order construct asymptotic confidence sets. Other examples are in Banerjee et al. (2016) who considered divide-and-conquer techniques for non-standard (i.e. non-"root-$n$") scalar estimators, and in Battey et al. (2015) and Tang et al. (2016), who both consider the use of divide-and-conquer techniques for regularized regression estimators such as the LASSO of Tibshirani (1996). Some non-asymptotic results for estimators obtained via divide-and-conquer optimization techniques are obtained in Zhang et al. (2013).

We proceed to present the vector extensions to the results of Matloff (2016c) and Li et al. (2013) in Section 2. Extensions to non-IID random sampling are presented in Section 3. Proofs of theoretical results are relegated to Section 4. Section 5 provides directions towards our supplementary materials, which present the results of simulation studies that lend empirical support to our theoretical results. The Appendix contains two interesting theorems that are required for our proofs.

## 2. Vector CA Estimators

Denote $\to$, $\xrightarrow{p}$, and $\rightsquigarrow$ as non-stochastic convergence, convergence in distribution, and convergence in distribution, respectively. Let $\boldsymbol{\mu}_n = \mathbb{E}\left(\hat{\boldsymbol{\theta}}_{jn}\right)$ and $\boldsymbol{\Sigma}_n = \mathrm{cov}\left(\hat{\boldsymbol{\theta}}_{jn}\right)$ be the mean vector and covariance matrix of $\hat{\boldsymbol{\theta}}_{jn}$, respectively. Assume that $\boldsymbol{\mu}_n$ and $\boldsymbol{\Sigma}_n$ exist. Write the multivariate normal distribution with mean $\boldsymbol{\mu}$ and covariance $\boldsymbol{\Sigma}$ as $N(\boldsymbol{\mu}, \boldsymbol{\Sigma})$ and let $\boldsymbol{Z}_j$ ($j \in \mathbb{N} \cup \{0\}$) be a random variable with distribution $N(\boldsymbol{0}, \boldsymbol{I})$, where $\boldsymbol{0}$ is a zero vector/matrix (in context) and $\boldsymbol{I}$ is an identity matrix. Denote any quantity that does not depend on an index with a subscript 0. Finally, let $\|\cdot\|$ denote the Euclidean norm and let the superscript $\top$ denote matrix transposition. We are now ready to present vector-parameter versions of the results from Matloff (2016c) and Li et al. (2013).

**Proposition 1.** *If $\{\boldsymbol{X}_i\}_{i=1}^n$ is IID, $\boldsymbol{\mu}_n \to \boldsymbol{\theta}$, and $c_n^{-1}\boldsymbol{\Sigma}_n \to \boldsymbol{0}$ (in $n$), then $\mathbb{E}\|\bar{\boldsymbol{\theta}}_n - \boldsymbol{\theta}\|^2 \to \boldsymbol{0}$.*

*Remark* 2. By application of the Markov inequality, Proposition 1 implies $\bar{\boldsymbol{\theta}}_n$ is a consistent estimator for $\boldsymbol{\theta}$. Furthermore, Proposition 1 is applicable in both the cases where $r_n$ is fixed and $c_n$ diverges to infinity, and where $r_n$ diverges and $c_n$ is fixed. For example, consider the cases where $\hat{\boldsymbol{\theta}}_{jn}$ is the sample mean vector obtained from either $r_n = r$ ($c_n \to \infty$) observations or $r_n = n$ ($c_n = 1$) observations, where the population that is sampled from has finite covariance matrix $\mathrm{cov}(\boldsymbol{X}_i) = \boldsymbol{\Sigma}$.

**Proposition 3.** *If $\{\boldsymbol{X}_i\}_{i=1}^n$ is IID, $c_n = c$ is fixed, and $\boldsymbol{\Sigma}_n^{-1/2}\left(\hat{\boldsymbol{\theta}}_{jn} - \boldsymbol{\mu}_n\right) \rightsquigarrow \boldsymbol{Z}_j$ for each $j \in [c]$, then*

$$c^{1/2}\boldsymbol{\Sigma}_n^{-1/2}\left(\bar{\boldsymbol{\theta}}_n - \boldsymbol{\mu}_n\right) \rightsquigarrow \boldsymbol{Z}_0.$$

*Remark* 4. As noted in the introduction, Proposition 3 was suggested in Matloff (2016a) and Matloff (2016b), but was not proved. Proposition 3 only applies when the $j$th chunk estimator $\hat{\boldsymbol{\theta}}_{jn}$ is already known to be asymptotically normal, as $r_n \to \infty$, and when $c_n = c$ is fixed at a constant value.

**Proposition 5.** *If $\{\boldsymbol{X}_i\}_{i=1}^n$ is IID, $r_n = r$ is fixed, then $\boldsymbol{\mu}_n = \boldsymbol{\mu}_0$ and $\boldsymbol{\Sigma}_n = \boldsymbol{\Sigma}_0$ are not dependent on $n$, and $c_n^{1/2}\boldsymbol{\Sigma}_0^{-1/2}\left(\bar{\boldsymbol{\theta}}_n - \boldsymbol{\mu}_0\right) \rightsquigarrow \boldsymbol{Z}_0$.*

*Proof.* The proposition is a simple result of the multivariate Lindberg-Lévy CLT (central limit theorem); see for example Boos and Stefanski (2013, Thm. 5.7). □

*Remark* 6. Proposition 5, note that $c_n^{1/2}\boldsymbol{\Sigma}_0^{-1/2}\left(\bar{\boldsymbol{\theta}}_n - \boldsymbol{\theta}\right) \rightsquigarrow \boldsymbol{Z}_0$ if and only if $\boldsymbol{\mu}_0 = \boldsymbol{\theta}$. If $\boldsymbol{\mu}_0 \neq \boldsymbol{\theta}$, then $\bar{\boldsymbol{\theta}}_n$ is inconsistent. This is because $r_n = r$ is fixed at a constant value, and thus $\boldsymbol{\mu}_n = \mathbb{E}\left(\hat{\boldsymbol{\theta}}_{jn}\right) = \boldsymbol{\mu}_0$ must also be a constant.

**Proposition 7.** *If $\{\boldsymbol{X}_i\}_{i=1}^n$ is IID and*

$$\mathbb{E}\left|\boldsymbol{t}^\top \hat{\boldsymbol{\theta}}_{jn} - \boldsymbol{t}^\top \boldsymbol{\mu}_n\right|^{2+\delta} / \left[c_n^{\delta/2}\left(\sqrt{\boldsymbol{t}^T \boldsymbol{\Sigma}_n \boldsymbol{t}}\right)^{2+\delta}\right] \to 0, \tag{2}$$

*as $n \to \infty$, for all $\boldsymbol{t} \in \mathbb{R}^d$ and some $\delta > 0$, then $c_n^{1/2}\boldsymbol{\Sigma}_n^{-1/2}\left(\bar{\boldsymbol{\theta}}_n - \boldsymbol{\mu}_n\right) \rightsquigarrow \boldsymbol{Z}_0$.*

*Remark* 8. Whereas the previous cases require either $c_n$ or $r_n$ to be fixed, Proposition 7 now allows for both to diverge to infinity, provided that condition (2) is met. Let $\hat{\theta}_{jkn}$ and $\mu_{kn}$ be the $k$th element of $\hat{\boldsymbol{\theta}}_{jn}$ and $\boldsymbol{\mu}_n$, for each $j \in [c_n]$ and $k \in [p]$, respectively. An alternative to condition (2) is to assume that $\mathbb{E}\left|\hat{\theta}_{jkn} - \mu_{kn}\right|^{2+\delta} < \Delta_1 < \infty$ for some $\delta > 0$, uniformly for all $j \in [c_n]$, $k \in [p]$, and $n$. This alternative assumption is stronger but easier to interpret than (2). We further note that Proposition 7 is the direct multivariate extension of Theorem 1(b) of Li et al. (2013).

*Remark* 9. By decomposition, we have

$$c_n^{1/2}\boldsymbol{\Sigma}_n^{-1/2}\left(\bar{\boldsymbol{\theta}}_n - \boldsymbol{\theta}\right) = c_n^{1/2}\boldsymbol{\Sigma}_n^{-1/2}\left(\bar{\boldsymbol{\theta}}_n - \boldsymbol{\mu}_n\right) + c_n^{1/2}\boldsymbol{\Sigma}_n^{-1/2}\left(\boldsymbol{\mu}_n - \boldsymbol{\theta}\right). \tag{3}$$

Thus, if we assume that $c_n^{1/2}\boldsymbol{\Sigma}_n^{-1/2}\left(\boldsymbol{\mu}_n - \boldsymbol{\theta}\right) \to \boldsymbol{0}$, then by a Slutsky-type theorem (cf. van der Vaart (1998, Thm. 2.7)) we have $c_n^{1/2}\boldsymbol{\Sigma}_n^{-1/2}\left(\bar{\boldsymbol{\theta}}_n - \boldsymbol{\theta}\right) \rightsquigarrow \boldsymbol{Z}_0$.

*Remark* 10. Let $\hat{\boldsymbol{\theta}}_n$ be the single-chunk estimator and suppose that $n^{1/2}\left(\hat{\boldsymbol{\theta}}_n - \boldsymbol{\theta}\right) \rightsquigarrow \boldsymbol{\Sigma}_0^{1/2}\boldsymbol{Z}_0$ for some positive definite covariance matrix $\boldsymbol{\Sigma}_0$. By subsampling, we also have $r_n^{1/2}\left(\hat{\boldsymbol{\theta}}_{jn} - \boldsymbol{\theta}\right) \rightsquigarrow \boldsymbol{\Sigma}_0^{1/2}\boldsymbol{Z}_j$, which implies, $n^{1/2}\left(\boldsymbol{\mu}_n - \boldsymbol{\theta}\right) \to \boldsymbol{0}$ (by a similar decomposition to (3)), $r_n^{1/2}\boldsymbol{\Sigma}_n^{1/2} \to \boldsymbol{\Sigma}_0^{1/2}$ (cf. Boos and Stefanski, 2013, Sec.

5.5.4), and the expansion

$$\begin{aligned} n^{1/2}\left(\bar{\boldsymbol{\theta}}_n - \boldsymbol{\theta}\right) &= n^{1/2}\left(\bar{\boldsymbol{\theta}}_n - \boldsymbol{\mu}_n\right) + n^{1/2}\left(\boldsymbol{\mu}_n - \boldsymbol{\theta}\right) \\ &= r_n^{1/2}\boldsymbol{\Sigma}_n^{1/2}c_n^{1/2}\boldsymbol{\Sigma}_n^{-1/2}\left(\bar{\boldsymbol{\theta}}_n - \boldsymbol{\mu}_n\right) + n^{1/2}\left(\boldsymbol{\mu}_n - \boldsymbol{\theta}\right). \end{aligned}$$

By Propositions 3, 5, or 7, we have $c_n^{1/2}\boldsymbol{\Sigma}_n^{-1/2}\left(\bar{\boldsymbol{\theta}}_n - \boldsymbol{\mu}_n\right) \rightsquigarrow \boldsymbol{Z}_0$. Upon application of Slutsky-type theorem, we have $n^{1/2}\left(\bar{\boldsymbol{\theta}}_n - \boldsymbol{\theta}\right) \rightsquigarrow \boldsymbol{\Sigma}_0^{1/2}\boldsymbol{Z}_0$. Thus, under the hypothesis of Proposition 1, the CA estimator is as efficient as the single-chunk estimator, which is estimated via the simultaneous use of all $n$ random variables.

Propositions 5 and 7 imply that one can conduct asymptotic inference for $\boldsymbol{\theta}$ using the random sample $\left\{\hat{\boldsymbol{\theta}}_{jn}\right\}_{j=1}^{c_n}$. Appealing to the multivariate Lindberg-Lévy CLT, we can approximate the distribution of $\bar{\boldsymbol{\theta}}_n$ by the multivariate normal distribution with mean $\boldsymbol{\theta}$ and estimated covariance $\bar{\boldsymbol{\Sigma}}_n = (c_n - p)^{-1} \sum_{j=1}^{c_n} \left(\hat{\boldsymbol{\theta}}_{jn} - \bar{\boldsymbol{\theta}}_n\right)\left(\hat{\boldsymbol{\theta}}_{jn} - \bar{\boldsymbol{\theta}}_n\right)^\top$. This allows for the construction of hypothesis tests and confidence intervals for $\boldsymbol{\theta}$ and its elements via conventional normal theory. The delta method can also be applied if one wishes to conduct inferences regarding some continuous function $\boldsymbol{h} : \mathbb{R}^p \to \mathbb{R}^q$ of $\boldsymbol{\theta}$ (cf. DasGupta, 2008, Sec. 3.4).

*Remark* 11. Unlike Propositions 5 and 7, the number of chunks $c_n = c$ is taken to be finite by the hypothesis of Proposition 3. As such, empirical methods can be used instead of appealing to the CLT. In Matloff (2016c), the Bootstrap is suggested for conducting inference.

## 3. Non-IID Results

Both Matloff (2016c) and Li et al. (2013) raised the issue of non-IID sampling as being important in the study of CA estimators. In Matloff (2016c), a remedy for a simple type of heterogeneity among the chunks is considered, whereupon the number of chunks $c_n = c$ is fixed and the $r_n$ random variables are homogeneous within each chunk, but heterogeneous between chunks. In such a scenario, one can randomly reassign the the random variables between chunks and thus break the heterogeneity. However, this is not always possible, especially when the chunks are stored at different locations. Our next proposition presents one alternative solution.

**Proposition 12.** *Let $c_n = c$ be fixed and let $\boldsymbol{X}_{1j}, ..., \boldsymbol{X}_{r_nj}$ be a random sample for each $j \in [c]$, such that $\boldsymbol{\mu}_{jn} = \mathbb{E}\left(\hat{\boldsymbol{\theta}}_{jn}\right)$ and $\boldsymbol{\Sigma}_n = \mathrm{cov}\left(\hat{\boldsymbol{\theta}}_{jn}\right)$. If $\boldsymbol{\Sigma}_n^{-1/2}\left(\hat{\boldsymbol{\theta}}_{jn} - \boldsymbol{\mu}_{jn}\right) \rightsquigarrow \boldsymbol{Z}_j$ for each $j$, then $c^{1/2}\boldsymbol{\Sigma}_n^{-1/2}\left(\bar{\boldsymbol{\theta}}_n - \boldsymbol{\mu}_n\right) \rightsquigarrow \boldsymbol{Z}_0$, where $\boldsymbol{\mu}_n = c^{-1}\sum_{j=1}^c \boldsymbol{\mu}_{jn}$.*

*Proof.* The proof is identical to that of Proposition 3, except one expands both $\bar{\boldsymbol{\theta}}_n$ and $\boldsymbol{\mu}_n$ in the LHS of the result. □

When $r_n$ and $c_n$ are both taken to increase with $n$, there are numerous modes of IID violations that can be considered. The double array Liapounov CLT yields the following result.

**Proposition 13.** *Let $\boldsymbol{X}_{1j}, ..., \boldsymbol{X}_{r_n j}$ be a random sample for each $j \in [c_n]$, such that such that $\boldsymbol{\mu}_{jn} = \mathbb{E}\left(\hat{\boldsymbol{\theta}}_{jn}\right)$ and $\boldsymbol{\Sigma}_{jn} = cov\left(\hat{\boldsymbol{\theta}}_{jn}\right)$. Assume that $\left\{\hat{\boldsymbol{\theta}}_{jn}\right\}_{j=1}^{c_n}$ is independently distributed with moments $\mathbb{E}\left|\hat{\theta}_{jkn} - \mu_{kn}\right|^{2+\delta} < \Delta_1 < \infty$ for some $\delta > 0$, uniformly for all $j \in [c_n]$, $k \in [p]$, and $n$. Define $\boldsymbol{\mu}_n = c_n^{-1}\sum_{j=1}^{c_n}\boldsymbol{\mu}_{jn}$ and $\boldsymbol{\Sigma}_n = c_n^{-1}\sum_{j=1}^{c_n}\boldsymbol{\Sigma}_{jn}$. If $\boldsymbol{\Sigma}_n$ is strictly positive definite for all sufficiently large $n$, then $c_n^{1/2}\boldsymbol{\Sigma}_n^{-1/2}\left(\bar{\boldsymbol{\theta}}_n - \boldsymbol{\mu}_n\right) \rightsquigarrow \boldsymbol{Z}_0$.*

*Remark* 14. Note that Proposition 13 is a generalization of Proposition 7, which utilizes an alternative form of the Liapounov CLT and that allows for heterogeneity in the mean vector and covariance matrix of each chunk. Further, Proposition 13 does not require the samples within each chunk to be IID.

We now consider correlations between chunks using the notions of mixingales and mixing processes. Theorems that are required to prove the subsequent results can be found in the Appendix.

**Definition 15** (White (2001, Def. 5.15)). *Let $\{X_j\}_{j=-\infty}^{\infty}$ be a random sequence with $\mathbb{E}\left(X_j^2\right) < \infty$ for all $j \in \mathbb{Z}$. Let $\mathcal{F}_j$ be a filtration that is adapted to the sequence $\{X_j\}_{j=-\infty}^{\infty}$. The sequence $\{X_j, \mathcal{F}_j\}_{j=-\infty}^{\infty}$ is an adapted mixingale if there exist finite nonnegative scalars $a_j$ and $b_l$ ($l \in \mathbb{N}$), such that $b_l \to 0$ as $l \to \infty$ and $\left(\mathbb{E}\left[\mathbb{E}\left(X_j|\mathcal{F}_{j-l}\right)^2\right]\right)^{1/2} \leq a_j b_l$. Further, we say that $b_l$ has size $-s$ if $b_l = O\left(1/l^{s+\epsilon}\right)$, for some $\epsilon > 0$.*

**Proposition 16.** *Let $r_n = r$ be fixed and $\boldsymbol{X}_{1j}, ..., \boldsymbol{X}_{rj}$ be a random sample for each $j \in [c_n]$, such that such that $\boldsymbol{\mu}_0 = \mathbb{E}\left(\hat{\boldsymbol{\theta}}_{jn}\right)$ and $\boldsymbol{\Sigma}_0 = cov\left(\hat{\boldsymbol{\theta}}_{jn}\right)$. Assume that $\left\{\hat{\boldsymbol{\theta}}_{jn}\right\}_{j=1}^{c_n}$ is a stationary ergodic sequence and that $\left\{\hat{\boldsymbol{\theta}}_{jn}, \mathcal{F}_j\right\}_{j=1}^{c_n}$ is an adapted mixingale of size $-1$ with moments $\mathbb{E}\left|\hat{\theta}_{jkn} - \mu_{kn}\right|^2 < \Delta_1 < \infty$, uniformly for all $j \in [c_n]$, $k \in [p]$, and $n$. Define $\boldsymbol{\Sigma}_n = c_n^{-1}cov\left(\hat{\boldsymbol{\theta}}_{jn} - \boldsymbol{\mu}_0\right)$. If $\boldsymbol{\Sigma}_n$ is strictly positive definite for all sufficiently large $n$, then $\boldsymbol{\Sigma}_n \to \boldsymbol{\Sigma}_0$ and $c_n^{1/2}\boldsymbol{\Sigma}_0^{-1/2}\left(\bar{\boldsymbol{\theta}}_n - \boldsymbol{\mu}_0\right) \rightsquigarrow \boldsymbol{Z}_0$.*

Let $\mathcal{B}_n^{m+n}$ denote the Borel $\sigma$-field generated by the sequence $\{\boldsymbol{X}_j\}_{j=n}^{m+n}$. If $\mathcal{A}$ and $\mathcal{B}$ are $\sigma$-fields, then we can define the functions

$$\phi(\mathcal{A}, \mathcal{B}) = \sup_{\{A \in \mathcal{A}, B \in \mathcal{B}: \mathbb{P}(A) > 0\}} |\mathbb{P}(B|A) - \mathbb{P}(B)|$$

and

$$\alpha(\mathcal{A}, \mathcal{B}) = \sup_{\{A \in \mathcal{A}, B \in \mathcal{B}\}} |\mathbb{P}(A \cap B) - \mathbb{P}(A)\mathbb{P}(B)|.$$

**Definition 17** (White (2001, Def. 3.16)). *Define $\phi(m) = \sup_n \phi\left(\mathcal{B}_{-\infty}^n, \mathcal{B}_{n+m}^\infty\right)$ and $\alpha(m) = \sup_n \alpha\left(\mathcal{B}_{-\infty}^n, \mathcal{B}_{n+m}^\infty\right)$. If $\phi(m) \to 0$ as $m \to \infty$, then $\{\boldsymbol{X}_j\}_{j=-\infty}^\infty$ is said to be $\phi$-mixing. If $\alpha(m) \to 0$ as $m \to \infty$, then $\{\boldsymbol{X}_j\}_{j=-\infty}^\infty$ is said to be $\alpha$-mixing or strongly mixing. Furthermore, if $\phi(m) = O\left(1/m^{s+\epsilon}\right)$ or $\alpha(m) = O\left(1/m^{s+\epsilon}\right)$, for some $\epsilon > 0$, then we say that the $\phi$- or $\alpha$-mixing coefficient of $\{\boldsymbol{X}_j\}_{j=-\infty}^\infty$ is of size $-s$, respectively.*

**Proposition 18.** *Let $X_{1j}, ..., X_{r_nj}$ be a random sample for each $j \in [c_n]$, such that such that $\boldsymbol{\mu}_{jn} = \mathbb{E}\left(\hat{\boldsymbol{\theta}}_{jn}\right)$ and $\boldsymbol{\Sigma}_{jn} = cov\left(\hat{\boldsymbol{\theta}}_{jn}\right)$. Assume that $\left\{\hat{\boldsymbol{\theta}}_{jn}\right\}_{j=1}^{c_n}$ is $\phi$-mixing with coefficient be of size $-r/2(r-1)$, or $\alpha$-mixing with coefficient be of size $-r/(r-2)$, with moments $\mathbb{E}\left|\hat{\theta}_{jkn} - \mu_{kn}\right|^r < \Delta_1 < \infty$ for some $r > 2$, uniformly for all $j \in [c_n]$, $k \in [p]$, and $n$. Define $\boldsymbol{\mu}_n = c_n^{-1} \sum_{j=1}^{c_n} \boldsymbol{\mu}_{jn}$ and $\boldsymbol{\Sigma}_n = c_n^{-1} \sum_{j=1}^{c_n} \boldsymbol{\Sigma}_{jn}$. If $\boldsymbol{\Sigma}_n$ is strictly positive definite for all sufficiently large $n$, then $c_n^{1/2} \boldsymbol{\Sigma}_n^{-1/2} \left(\bar{\boldsymbol{\theta}}_n - \boldsymbol{\mu}_n\right) \rightsquigarrow \boldsymbol{Z}_0$.*

*Remark* 19. In Li et al. (2013), it was proposed that one may analyze the CA estimator under the condition that $\{X_j\}_{j=1}^n$ is $M$-dependent. That is each $X_j$ is independent of all but at most $M < \infty$ other random variables $X_k$, such that $j \neq k$. If we assume that the random samples within each chunk $X_{1j}, ..., X_{r_nj}$ be a random sample for each $j \in [c_n]$, but not between, then any of the Propositions 12, 13, 16, and 18 are applicable, given the further assumptions of their hypotheses. This is because the $M$-dependence is contained within each chunk and the between chunk dependences are not affected. However, if the entire data is $M$-dependent, then we may still utilize Proposition 18. This is because $M$-dependence between chunks clearly implies $M$-dependence within chunks. Furthermore, $M$-dependence implies both $\phi$- and $\alpha$-mixing, and thus allows for the satisfaction of the hypothesis for Proposition 18 (cf. Bradley, 2005).

*Remark* 20. The mixing of Definition 17 only allows for serially-dependent observations. If one requires spatial, spatial-temporal, or other modes of dependence, then alternative definitions are required. A generic law of large numbers and CLT for lattice-based mixing that can be used to generalize Proposition 18 is available from Jenish and Prucha (2009). It is noted in Bradley (1989) that $M$-dependence is equivalent to both $\phi$ and $\alpha$-mixing on lattice random fields.

*Remark* 21. Theorem 5.17 of Boos and Stefanski (2013) states that if $c_n^{1/2} \boldsymbol{\Sigma}_n^{-1/2} \left(\bar{\boldsymbol{\theta}}_n - \boldsymbol{\mu}_n\right) \rightsquigarrow \boldsymbol{Z}_0$ and $c_n^{-1/2} \boldsymbol{\Sigma}_n \to \boldsymbol{0}$, then $\bar{\boldsymbol{\theta}}_n \xrightarrow{p} \boldsymbol{\mu}_n$. However, this does not guarantee consistency (i.e. $\bar{\boldsymbol{\theta}}_n \xrightarrow{p} \boldsymbol{\theta}$) as $\boldsymbol{\mu}_n$ may not approach $\boldsymbol{\theta}$ in the limit. In order to obtain consistency, we can perform a decomposition similar to (3), which reveals the requirement that $c_n^{1/2} \boldsymbol{\Sigma}_n^{-1/2} \left(\boldsymbol{\mu}_n - \boldsymbol{\theta}\right) \to \boldsymbol{0}$ for obtaining $c_n^{1/2} \boldsymbol{\Sigma}_n^{-1/2} \left(\bar{\boldsymbol{\theta}}_n - \boldsymbol{\theta}\right) \rightsquigarrow \boldsymbol{Z}_0$, in Propositions 12, 13, and 18. We can then make the additional assumption that $c_n^{-1/2} \boldsymbol{\Sigma}_n \to \boldsymbol{0}$ in order to obtain $\bar{\boldsymbol{\theta}}_n \xrightarrow{p} \boldsymbol{\theta}$. A similar conclusion can be made regarding Proposition 16 if we assume that $\boldsymbol{\mu}_0 = \boldsymbol{\theta}$.

*Remark* 22. Although Propositions 13 and 18 are stated with the allowance for $r_n$ to increase as $n$ increases, the same results hold for fixed $r_n = r$. Remark 6 then applies when establishing the consistency of the CA estimator $\bar{\boldsymbol{\theta}}_n$.

*Remark* 23. Similarly to Remark 10, we can show that the CA estimator $\bar{\boldsymbol{\theta}}_n$ is equally as efficient to the single chunk estimator $\hat{\boldsymbol{\theta}}_n$ in the non-IID cases. Assume that the sample $\{X_i\}_{i=1}^n$ arises as a subsequence of a stationary random process that satisfies either the hypotheses of Proposition 16 or 18, and suppose that each of the $c_n$ chunks of observations contains $r_n$ contiguous observations from the sample. Under the assumption that $n^{1/2}\left(\hat{\boldsymbol{\theta}}_n - \boldsymbol{\theta}\right) \rightsquigarrow \boldsymbol{\Sigma}_0^{1/2} \boldsymbol{Z}_0$, for some positive definite covariance matrix $\boldsymbol{\Sigma}_0$, we have the same conclusion regarding relative efficiency as that of Remark 10. The additional assumptions required to establish the same result under the hypotheses of Propositions 12 and 13 make the cases equivalent to the hypotheses of Propositions 3 and 7, respectively.

## 4. Proofs

*Proof of Proposition 1*

By expansion, $\mathbb{E} \left\|\bar{\boldsymbol{\theta}}_n - \boldsymbol{\theta}\right\|^2 = \mathbb{E} \left\|\bar{\boldsymbol{\theta}}_n\right\|^2 - 2\boldsymbol{\theta}^\top \mathbb{E}\left(\bar{\boldsymbol{\theta}}_n\right) + \mathbb{E} \left\|\bar{\boldsymbol{\theta}}_n\right\|^2$. Applying the fact $\mathbb{E}\left(\bar{\boldsymbol{\theta}}_n\right) = \boldsymbol{\mu}_n$ and Petersen and Pedersen (2008, Eqn. 296), yields $\mathbb{E} \left\|\bar{\boldsymbol{\theta}}_n - \boldsymbol{\theta}\right\|^2 = \text{tr}\left(c_n^{-1} \boldsymbol{\Sigma}_n\right) + \|\boldsymbol{\mu}_n\|^2 - 2\boldsymbol{\theta}^\top \boldsymbol{\mu}_n + \|\boldsymbol{\theta}\|^2$. Lastly, by the hypothesis, we have $\mathbb{E} \left\|\bar{\boldsymbol{\theta}}_n - \boldsymbol{\theta}\right\|^2 \to \mathbf{0}$.

*Proof of Proposition 3*

By expansion, the left-hand side (LHS) of the result is $c^{-1/2} \sum_{j=1}^c \boldsymbol{\Sigma}_n^{-1/2}\left(\hat{\boldsymbol{\theta}}_{jn} - \boldsymbol{\mu}_n\right)$ thus LHS $\rightsquigarrow c^{-1/2} \sum_{j=1}^c \boldsymbol{Z}_j$. Next, $c^{-1/2} \sum_{j=1}^c \boldsymbol{Z}_j = \boldsymbol{Z}_0$ by the additive closure of normal random variables (e.g. Seber, 2008, Eqn. 20.23).

*Proof of Proposition 7*

We have
$$\frac{\sum_{i=1}^{c_n} \mathbb{E}\left|\boldsymbol{t}^\top \hat{\boldsymbol{\theta}}_{jn} - \boldsymbol{t}^\top \boldsymbol{\mu}_n\right|^{2+\delta}}{\left(\sum_{i=1}^{c_n} \boldsymbol{t}^\top \boldsymbol{\Sigma}_n \boldsymbol{t}\right)^{(2+\delta)/2}} = \frac{1}{c_n^{\delta/2}} \mathbb{E}\left|\frac{\boldsymbol{t}^\top \hat{\boldsymbol{\theta}}_{1n} - \boldsymbol{t}^\top \boldsymbol{\mu}_n}{\sqrt{\boldsymbol{t}^\top \boldsymbol{\Sigma}_n \boldsymbol{t}}}\right|^{2+\delta} \to 0,$$
in $n$, as $c_n \to \infty$ by the hypothesis. Further, the Liapounov CLT conditions hold (e.g. DasGupta (2008, Thm. 5.2)) and therefore $c_n^{1/2} \sum_{j=1}^c \left(\boldsymbol{t}^T \boldsymbol{\Sigma}_n \boldsymbol{t}\right)^{-1} \left(\boldsymbol{t}^\top \bar{\boldsymbol{\theta}}_n - \boldsymbol{t}^\top \boldsymbol{\mu}_n\right) \rightsquigarrow Z_0$, for all $\boldsymbol{t} \in \mathbb{R}^d$, by the CLT. The desired result is obtained via the Cramér-Wold device (cf. van der Vaart, 1998, Example 2.18).

*Proof of Proposition 13*

Following the proof of White (2001, Thm. 5.13), we wish to show that $c_n^{1/2} \boldsymbol{t}^\top \boldsymbol{\Sigma}_n^{-1/2} \left(\hat{\boldsymbol{\theta}}_{jn} - \boldsymbol{\mu}_n\right) \rightsquigarrow \boldsymbol{Z}_0$ for all $\boldsymbol{t} \in \mathbb{R}^p$, using the double array Liapounov CLT (e.g. White, 2001, Thm 5.11). This requires that $\left\{\boldsymbol{t}^\top \hat{\boldsymbol{\theta}}_{jn}\right\}_{j=1}^n$ be IID and that $\mathbb{E}\left|\boldsymbol{t}^\top \left(\hat{\boldsymbol{\theta}}_{jn} - \boldsymbol{\mu}_n\right)\right|^{2+\delta} < \Delta_1 < \infty$ for some $\delta > 0$, for all $\boldsymbol{t}$, $j$ and $n$. These conditions are fulfilled by the assumptions that $\left\{\hat{\boldsymbol{\theta}}_{jn}\right\}_{j=1}^n$ be IID, and $\mathbb{E}\left|\hat{\theta}_{jkn} - \mu_{kn}\right|^{2+\delta} < \Delta_2 < \infty$ for all $j$, $k$, and $n$, respectively. Further, it must be assumed that $\boldsymbol{t}^\top \boldsymbol{\Sigma}_n \boldsymbol{t} > 0$ for all $\boldsymbol{t}$. This is the definition of strict positive definiteness (cf. Seber, 2008, Def. 10.1). The result then follows via an application of the Cramér-Wold device.

*Proof of Proposition 16*

Following the proof of White (2001, Thm. 5.17), we wish to show that $c_n^{1/2} t^\top \Sigma_0^{-1/2} \left( \hat{\theta}_{jn} - \mu_0 \right) \rightsquigarrow Z_0$ for all $t \in \mathbb{R}^p$ and some fixed $\Sigma_0$, using Theorem 24. First, set $X_j = t^\top \Sigma_0^{-1/2} \left( \hat{\theta}_{jn} - \mu_0 \right)$ and note that it is stationary ergodic under the hypothesis. To check that $\mathbb{E}\left(X_j^2\right) < \infty$, we note that we can write $X_j = \sum_{k=1}^p \tilde{t}_k \left( \hat{\theta}_{jkn} - \mu_{k0} \right)$, where $\tilde{t}_k$ is the $k$th element of $\tilde{t} = \Sigma_0^{-1/2} t$. By definition of $t$ and $\Sigma_n$, we have $|\tilde{t}_k| < \Delta_2 < \infty$, uniformly for all $k$. By Minkowski's inequality, we have

$$\mathbb{E}\left(X_j^2\right) \leq \left[ \Delta_2 \sum_{k=1}^p \left( \mathbb{E} \left| \tilde{t}_k \left( \hat{\theta}_{jkn} - \mu_{k0} \right) \right|^2 \right)^{1/2} \right]^2$$

$$\leq \left[ \Delta_2 \sum_{k=1}^p \mathbb{E} \left| \hat{\theta}_{jkn} - \mu_{k0} \right| \right]^2 \leq \left( \Delta_2 k \Delta_1^{1/2} \right)^2 < \infty$$

under the assumption that $\mathbb{E} \left| \hat{\theta}_{jkn} - \mu_{k0} \right|^2 < \Delta_1$ and stationarity. Next, we must demonstrate that $\{X_j \mathcal{F}_j\}_{j=-\infty}^\infty$ is a mixingale of size $-1$. We write

$$\mathbb{E}\left[ \mathbb{E}\left( X_0 | \mathcal{F}_{-l} \right)^2 \right] = \mathbb{E}\left[ \mathbb{E}\left( \sum_{k=1}^p \tilde{t}_k \left( \hat{\theta}_{0kn} - \mu_{k0} \right) | \mathcal{F}_{-l} \right)^2 \right].$$

Again, by Minkowski's inequality, we have

$$\mathbb{E}\left[ \mathbb{E}\left( X_0 | \mathcal{F}_{-l} \right)^2 \right]$$

$$\leq \left[ \Delta_2 \sum_{k=1}^p \left( \mathbb{E}\left[ \mathbb{E}\left( \sum_{k=1}^p \tilde{t}_k \left( \hat{\theta}_{0kn} - \mu_{k0} \right) | \mathcal{F}_{-l} \right)^2 \right] \right)^{1/2} \right]^2$$

$$\leq \left( \Delta_2 \sum_{k=1}^p a_{0k} b_{kl} \right)^2 \leq \left( \Delta_2 p \bar{a}_{0k} \bar{b}_{kl} \right)^2,$$

where $\bar{a}_0 = \max_k \bar{a}_{0k}$, and $\bar{b}_l = \max_k b_{kl}$ is of size $-1$, by the assumption that $\left\{ \hat{\theta}_{jn}, \mathcal{F}_j \right\}_{j=1}^{c_n}$ is a mixingale of size $-1$. Lastly, we have

$$c_n^{-1} \mathrm{var}\left( \sum_{j=1}^{c_n} X_j \right) = \mathrm{var}\left[ t^\top \Sigma_0^{-1/2} \left( \hat{\theta}_{jn} - \mu_0 \right) \right]$$

$$= t^\top \Sigma_0^{-1/2} \mathrm{cov}\left( \hat{\theta}_{jn} - \mu_0 \right) \Sigma_0^{-1/2} t$$

$$= t^\top \Sigma_0^{-1/2} \Sigma_n \Sigma_0^{-1/2} t \to \bar{\sigma}_0^2 < \infty$$

as $n \to \infty$, for all $t$. Thus, $\Sigma_n$ converges to a finite limit. We now set $\Sigma_0 = \lim_{n \to \infty} \Sigma_n$ and obtain the desired result via an application of the Cramér-Wold device.

*Proof of Proposition 18*

Following the proof of White (2001, Thm. 5.23), we wish to show that $c_n^{1/2} t^\top \Sigma_n^{-1/2} (\hat{\boldsymbol{\theta}}_{jn} - \boldsymbol{\mu}_n) \rightsquigarrow \boldsymbol{Z}_0$ for all $t \in \mathbb{R}^p$, using Theorem 25. Since $\{\hat{\boldsymbol{\theta}}_{jn}\}_{j=1}^{c_n}$ is $\phi$- or $\alpha$-mixing with the hypothesized rates, so is $X_{jn} = t^\top \Sigma_n^{-1/2} (\hat{\boldsymbol{\theta}}_{jn} - \boldsymbol{\mu}_n)$ by the continuous mapping theorem of White (2001, Thm. 3.49). Next, $\mathbb{E}(X_{jn}) = 0$ by definition of the mean and $\mathbb{E}|X_{jn}|^r < \Delta_2 < \infty$, for all $t$, by the assumption that $\mathbb{E}|\hat{\theta}_{jkn} - \mu_{kn}|^r < \Delta_1$. Finally,

$$\bar{\sigma}_n^2 = c_n^{-1} \text{var}\left(\sum_{j=1}^{c_n} X_{jn}\right) = t^\top \Sigma_n^{-1/2} \Sigma_n \Sigma_n^{-1/2} t = t^\top t < \infty$$

for all $t$, since $\Sigma_n$ is strictly positive definite. We obtain the desired result via an application of the Cramér-Wold device.

## 5. Simulation Studies

A comprehensive set of simulation studies are used to examine the behavior and performance of the CA estimator $\bar{\boldsymbol{\theta}}_n$ in comparison to the single chunk estimator $\hat{\boldsymbol{\theta}}_n$, across a range of estimation problems. The studies examine both the comparative accuracies of the estimators as well as the computational efficiencies of either approach. The results of the simulation studies can be found in the supplementary materials, available at http://tinyurl.com/SPL-SM-NM-2017.

## Acknowledgements

We are grateful to the reviewers, as their suggestions have led to great improvements to our paper. The authors are both funded by Australian Research Council grants.

## Appendix

The following two theorems are used in the proofs of Propositions 16 and 18, respectively.

**Theorem 24** (White (2001, Thm. 5.16)). *If $\{X_j, \mathcal{F}_j\}_{j=-\infty}^{\infty}$ is a stationary ergodic adapted mixingale of size $-1$, then $\bar{\sigma}_n^2 = n^{-1} \text{var}\left(\sum_{j=1}^n X_j\right) \to \bar{\sigma}_0^2$ as $n \to \infty$. Further, if $\sigma_0^2 > 0$, then $n^{-1/2} \bar{X}_j / \bar{\sigma}_0 \rightsquigarrow Z_0$.*

**Theorem 25** (White (2001, Thm. 5.20)). *Let $\{X_{jn}\}_{j=1}^{n}$ be a double array with mean $\mu_{jn} = \mathbb{E}(X_{jn}) = 0$ and variance $\sigma_{jn}^2 = var(X_{jn})$, such that $\mathbb{E}|X_{jn}|^r < \Delta_1 < \infty$ for some $r > 2$, uniformly for all $j$ and $n$. Further, let the $\phi$-mixing coefficient be of size $-r/2(r-1)$, or the $\alpha$-mixing coefficient be of size $-r/(r-2)$. If $\bar{\sigma}_n^2 = n^{-1} var\left(\sum_{j=1}^n X_{jn}\right) > 0$ for all sufficiently large $n$, then $n^{1/2}\left(\bar{X}_n - \bar{\mu}_n\right)/\bar{\sigma}_n \rightsquigarrow Z_0$, where $\bar{X}_n = n^{-1}\sum_{j=1}^n X_{jn}$.*